\documentstyle[sprocl]{article}

\input epsf 

\def\be{\begin{equation}}
\def\ee{\end{equation}}
\def\bea{\begin{eqnarray}}
\def\eea{\end{eqnarray}}

\def\msun{M_{\odot}} 
\def\mpl{m_{{\rm Pl}}}
\def\delh{\delta_{{\rm H}}}
\def\x{{\bf x}} 
\def\k{{\bf k}} 
 
\def\0{{\bf 0}} 
\def\1{{\bf 1}} 
\def\2{{\bf 2}}

\begin{document} 

\title{AN INTRODUCTION TO COSMOLOGICAL INFLATION}
\author{ANDREW R.~LIDDLE}
\address{Astronomy Centre, University of Sussex, Brighton BN1 9QJ,
U.~K.\\
and\\
Astrophysics Group, The Blackett Laboratory, Imperial College,\\ London
SW7 2BZ, U.~K. (present address)}

\maketitle
\abstracts{An introductory account is given of the inflationary
cosmology, which postulates a period of accelerated 
expansion during the Universe's earliest stages. The historical
motivation is briefly outlined, and the modelling of the inflationary
epoch explained. The most important aspect of inflation is that it
provides a possible model for the origin of structure in the Universe,
and key results are reviewed, along with a discussion of the current
observational situation and outlook.}
 

\section{Overview} 
 
One of the central planks of modern cosmology is the idea of inflation.
Originally introduced by Guth~\cite{Guth} in order to explain the
initial conditions for the hot big bang model, it has subsequently been
given a much more important role as the currently-favoured candidate for
the origin of structure in the Universe, such as galaxies, galaxy
clusters and cosmic microwave background anisotropies. This article
seeks to give an introductory account of the inflationary cosmology,
with the focus aimed towards inflation as a model for the origin of
structure. 

It begins with a quick review of the big bang cosmology, and the
problems with it which led to the introduction of inflation. The
modelling of the inflationary epoch using scalar fields is described,
and then results giving the form of perturbations produced by inflation
are quoted. Finally, the current observational situation is briefly
sketched. 

\section{Big bang problems and the idea of inflation}

The standard hot big bang theory is an extremely successful one, passing 
some crucial observational tests of which I'd highlight five.
\begin{itemize}
\item The expansion of the Universe.
\item The existence and spectrum of the cosmic microwave background 
radiation.
\item The abundances of light elements in the Universe (nucleosynthesis).
\item That the predicted age of the Universe is comparable to direct age 
measurements of objects within the Universe.
\item That {\em given} the irregularities seen in the microwave background 
by COBE, there exists a reasonable explanation for the development of 
structure in the Universe, through gravitational collapse.
\end{itemize}
In combination, these are extremely compelling. However, the standard
hot  big bang theory is limited to those epochs where the Universe is
cool enough  that the underlying physical processes are well established
and understood  through terrestrial experiment. It does not attempt to
address the state of  the Universe at earlier, hotter, times.
Furthermore, the hot big bang theory  leaves a range of crucial
questions unanswered, for it turns out that it can  successfully proceed
only if the initial conditions are very carefully  chosen. The
assumption of early Universe studies is that the mysteries of  the
conditions under which the big bang theory operates may be explained 
through the physics occurring in its distant, unexplored past. If so, 
accurate observations of the present state of the Universe may highlight
the  types of process occurring during these early stages, and perhaps
even shed  light on the nature of physical laws at energies which it
would be  inconceivable to explore by other means.

\subsection{A hot big bang reminder}

To get us started, I'll give a quick review of the big bang cosmology.
More detailed accounts can be found in any of a number of cosmological
textbooks. One of my aims in this section is to set down the notation
for the rest of the article.

\subsection{Equations of motion}

The hot big bang theory is based on the {\em cosmological principle}, which 
states that the Universe should look the same to all observers. That tells 
us that the Universe must be homogeneous and isotropic, which in turn tells 
us which metric must be used to describe it. It is the Robertson--Walker 
metric
\begin{equation}
ds^2 = -dt^2 + a^2(t) \left[ \frac{dr^2}{1-kr^2} + r^2 \left( d\theta^2
	+ \sin^2 \theta \, d\phi^2 \right) \right] \,.
\end{equation}
Here $t$ is the time variable, and $r$--$\theta$--$\phi$ are (polar) 
coordinates. The constant $k$ measures the spatial curvature, with $k$ 
negative, zero and positive corresponding to open, flat and closed
Universes  respectively. If $k$ is zero or negative, then the range of
$r$ is from zero  to infinity and the Universe is infinite, while if $k$
is positive then $r$  goes from zero to $1/\sqrt{k}$. Usually the
coordinates are rescaled to make  $k$ equal to $-1$, $0$ or $+1$. The
quantity $a(t)$ is the scale-factor of  the Universe, which measures its
physical size. The form of $a(t)$  depends on the properties of the
material within the Universe, as we'll see.

If no external forces are acting, then a particle at rest at a given set of 
coordinates $(r,\theta,\phi)$ will remain there. Such coordinates are said 
to be {\em comoving} with the expansion. One swaps between physical (ie 
actual) and comoving distances via
\begin{equation}
\mbox{physical distance} = a(t) \times \mbox{comoving distance} \,.
\end{equation}

The expansion of the Universe is governed by the properties of material 
within it. This can be specified~\footnote{I follow standard cosmological 
practice of setting the fundamental constants $c$ and $\hbar$ equal to one. 
This makes the energy density and mass density interchangeable (since the 
former is $c^2$ times the latter). I shall also normally use the Planck mass 
$\mpl$ rather than the gravitational constant $G$; with the convention just 
mentioned they are related by $G \equiv \mpl^{-2}$.} by the energy density 
$\rho(t)$ and the pressure $p(t)$. These are often related by an equation of 
state, which gives $p$ as a function of $\rho$; the classic examples are
\begin{eqnarray}
p & = & \frac{\rho}{3} \quad \quad \mbox{Radiation} \,, \\
p & = & 0 \quad \quad \; \; \mbox{Non-relativistic matter} \,.
\end{eqnarray}
In general though there need not be a simple equation of state; for example 
there may be more than one type of material, such as a combination of 
radiation and non-relativistic matter, and certain types of material, such 
as a scalar field (a type of material we'll encounter later which is 
crucial for modelling inflation), 
cannot be described by an equation of state at all.

The crucial equations describing the expansion of the Universe are
\begin{eqnarray}
H^2 = \frac{8\pi}{3 \mpl^2} \, \rho - \frac{k}{a^2} \quad \quad & &
	\mbox{Friedmann equation} \\
\dot{\rho} + 3 H (\rho+p) = 0  \quad \quad & &
	\mbox{Fluid equation}
\end{eqnarray}
where overdots are time derivatives and $H = \dot{a}/a$ is the Hubble 
parameter. The terms in the fluid equation contributing to $\dot{\rho}$ have 
a simple interpretation; the term $3H\rho$ is the reduction in density due 
to the increase in volume, and the term $3Hp$ is the reduction in energy 
caused by the thermodynamic work done by the pressure when this expansion 
occurs.

These can also be combined to form a new equation
\begin{equation}
\frac{\ddot{a}}{a} = - \frac{4\pi}{3\mpl^2} \left( \rho + 3p \right) \quad
	\quad \mbox{Acceleration equation} 
\end{equation}
in which $k$ does not appear explicitly.

\subsection{Standard cosmological solutions}

When $k= 0$ the Friedmann and fluid equations can readily be solved for
the equations of state given earlier, leading to the classic
cosmological solutions
\begin{eqnarray}
&\mbox{Matter Domination~~~~~~} p = 0 : & \rho \propto a^{-3} 
	\quad \quad a(t) \propto t^{2/3} \\
&\mbox{Radiation Domination~~~~~~} p = \rho/3 : & \rho \propto a^{-4} 
	\quad \quad a(t) \propto t^{1/2} 
\end{eqnarray}
In both cases the density falls as $t^{-2}$. When $k=0$ we have the freedom
to rescale $a$ and it is normally chosen to be unity at the present,
making physical and comoving scales coincide. The proportionality
constants are then fixed by setting the density to be $\rho_0$ at time
$t_0$, where here and throughout the subscript zero indicates present
value. 

A more intriguing solution appears for the case of a so-called
cosmological constant, which corresponds to an equation of state $p =
-\rho$. The fluid equation then gives $\dot{\rho} = 0$ and hence $\rho =
\rho_0$, leading to 
\begin{equation}
a(t) \propto \exp \left( Ht \right) \,.
\end{equation}

More complicated solutions can also be found for mixtures of components.
For example, if there is both matter and radiation the Friedmann
equation can be solved be using conformal time $\tau = \int dt/a$, while
if there is matter and a non-zero curvature term the solution can be
given either in parametric form using normal time $t$, or in closed form
with conformal time.

\subsection{Critical density and the density parameter}

The spatial geometry is flat if $k = 0$. For a given $H$, this requires that 
the density equals the critical density
\begin{equation}
\rho_{{\rm c}}(t) = \frac{3 \mpl^2 H^2}{8\pi} \,.
\end{equation}
Densities are often measured as fractions of $\rho_{{\rm c}}$:
\begin{equation}
\Omega(t) \equiv \frac{\rho}{\rho_{{\rm c}}} \,.
\end{equation}
The quantity $\Omega$ is known as the density parameter, and can be
applied to individual types of material as well as the total density.

The present value of the Hubble parameter is still not that well known, and 
is normally parametrized as
\begin{equation}
H_0 = 100 h \; {\rm km \, s}^{-1} \, {\rm Mpc}^{-1} = \frac{h}{3000} 
	\, {\rm Mpc}^{-1} \,,
\end{equation}
where $h$ is normally assumed to lie in the range $0.5 \leq h \leq 0.8$. The 
present critical density is
\begin{equation}
\rho_{{\rm c}}(t_0) = 1.88 \, h^2 \times 10^{-29} \, {\rm g \, cm}^{-3} =
	2.77 \, h^{-1} \times 10^{11} \, \msun/(h^{-1} {\rm Mpc})^3 \,.
\end{equation}

\subsection{Characteristic scales and horizons}

The big bang Universe has two characteristic scales
\begin{itemize}
\item The Hubble time (or length) $H^{-1}$.
\item The curvature scale $a|k|^{-1/2}$.
\end{itemize}
The first of these gives the characteristic timescale of evolution of 
$a(t)$, and the second gives the distance up to which space can be taken as
having a
flat (Euclidean) geometry. 
As written above they are both physical scales; to obtain the 
corresponding 
comoving scale one should divide by $a(t)$. The ratio of these scales 
actually gives a measure of $\Omega$; from the Friedmann equation we find
\begin{equation}
\sqrt{|\Omega -1|} = \frac{H^{-1}}{a |k|^{-1/2}} \,.
\end{equation}

A crucial property of the big bang Universe is that it possesses {\em 
horizons}; even light can only have travelled 
a finite distance since the start of 
the Universe $t_*$, given by
\begin{equation}
d_{{\rm H}}(t) = a(t) \int_{t_*}^{t} \frac{dt}{a(t)} \,.
\end{equation}
For example, matter domination gives $d_{{\rm H}}(t) = 3t = 2H^{-1}$. 
In a big bang Universe, $d_{{\rm H}}(t_0)$ is a good approximation to the 
distance to the surface of last scattering (the origin of the observed
microwave background, at a time known as `decoupling'), 
since $t_0 \gg t_{{\rm dec}}$.

\subsection{Redshift and temperature}

The redshift measures the expansion of the Universe via the stretching
of light
\begin{equation}
1+z = \frac{a(t_0)}{a(t_{{\rm emission}})} \,.
\end{equation}
Redshift can be used to describe both time and distance. As a time, it
simply refers to the time at which light would have to be emitted to have
a present redshift $z$. As a distance, it refers to the {\em present}
distance to an object from which light is received with a redshift $z$.
Note that this distance is not necessarily the time multiplied by the
speed of light, since the Universe is expanding as the light travels
across it.

As the Universe expands, it cools according to the law
\begin{equation}
T \propto \frac{1}{a} \,.
\end{equation}
In its earliest stages the Universe may have been arbitrarily hot and
dense. 

\subsection{The history of the Universe}

Presently the Universe is dominated by non-relativistic matter, but
because radiation reduces more quickly with the expansion, this implies
that at earlier times the Universe was radiation dominated. During the
radiation era temperature and time are related by
\begin{equation}
\frac{t}{1 \, {\rm sec}} \simeq 
    \left(\frac{10^{10} \, {\rm K}}{T} \right)^2 \,.
\end{equation}
The highest energies accessible to terrestrial experiment, generated
in particle accelerators, correspond to a temperature of about $10^{15}
\, {\rm K}$, which was attained when the Universe was about $10^{-10} \,
{\rm sec}$ old. Before that, we have no direct evidence of the
applicable physical laws and must use extrapolation based on current
particle physics model building. After that time there is a fairly clear
picture of how the Universe evolved to reach the present, with the key
events being as follows:
\begin{itemize}
\item $10^{-4}$ seconds: Quarks condense to form protons and neutrons.
\item 1 second: The Universe has cooled sufficiently that light nuclei
are able to form, via a process known as {\bf nucleosynthesis}.
\item $10^4$ years: The radiation density drops to the level of the
matter density, the epoch being known as {\bf matter--radiation
equality}. Subsequently the Universe is matter dominated.
\item $10^5$ years: Decoupling of radiation from matter leads to the
formation of the microwave background. This is more or less coincident
with recombination, when the up-to-now free electrons combine with the
nuclei to form atoms.
\item $10^{10}$ years: The present.
\end{itemize}

\section{Problems with the Big Bang}

In this section I shall quickly review the original motivation for the 
inflationary cosmology. These problems were largely ones of initial 
conditions. While historically these problems were very important, they
are  now somewhat marginalized as focus is instead concentrated on
inflation as a  theory for the origin of cosmic structure. 

\subsection{The flatness problem}

Taking advantage of the definition of the density parameter, and
ignoring a possible cosmological constant contribution, the 
Friedmann equation can be written in the form
\begin{equation}
|\Omega-1| = \frac{|k|}{a^2 H^2} \,.
\end{equation}
During standard big bang evolution, $a^2 H^2$ is decreasing, and so $\Omega$ 
moves away from one, for example
\begin{eqnarray}
& \mbox{Matter domination:} & |\Omega-1| \propto t^{2/3} \\
& \mbox{Radiation domination:} & |\Omega-1| \propto t
\end{eqnarray}
where the solutions apply provided $\Omega$ is close to one. So $\Omega = 1$ 
is an {\em unstable} critical point. Since we know that today $\Omega$ is 
certainly within an order of magnitude of one, it must have been much closer 
in the past. Inserting the appropriate behaviours for the matter and
radiation eras (or if you like just assuming radiation domination all
the way to the present) gives
\begin{eqnarray}
& \mbox{nucleosynthesis ($t \sim 1 \, {\rm sec}$)} : & |\Omega-1| < 
	{\cal O}(10^{-16}) \\
& \mbox{electro-weak scale ($t \sim 10^{-11} \, {\rm sec}$)} : & 
	|\Omega-1| < {\cal O}(10^{-27})
\end{eqnarray}
That is, hardly any choices of the initial density lead to a Universe like 
our own. Typically, the Universe will either swiftly recollapse, or will 
rapidly expand and cool below 3K within its first second of existence.

\subsection{The horizon problem}

Microwave photons emitted from opposite sides of the sky appear to be in 
thermal equilibrium at almost the same temperature. The most natural
explanation for this is that the Universe has indeed reached a state of
thermal equilibrium, through interactions between the different regions.
But unfortunately in the big bang theory this is not possible.
There was no time 
for those regions to interact before the photons were emitted, because of 
the finite horizon size,
\begin{equation}
\int_{t_*}^{t_{{\rm dec}}} \frac{dt}{a(t)} \ll 
	\int^{t_0}_{t_{{\rm dec}}} \frac{dt}{a(t)} \,. 
\end{equation}
This says that the distance light could travel before the microwave
background was released is much smaller than the present horizon distance.
In fact, any regions separated by more than about 2 degrees would be 
causally separated at decoupling in the hot big bang theory. In the big
bang theory there is therefore no explanation of why the Universe
appears so homogeneous.

In more recent years this problem has been brought into sharper focus
through the improving understanding of irregularities in the Universe,
as will be discussed later in this article. The same argument that
prevents the smoothing of the Universe also prevents the creation of
irregularities. For example, as we will see the COBE satellite observes
irregularities on all accessible angular scales, from a few degrees
upwards. In the simplest cosmological models, where these
irregularities are intrinsic to the last scattering surface, the 
perturbations are on too large a scale to have been created between the
big  bang and the time of decoupling, because the horizon size at
decoupling  subtends only a degree or so. Hence these perturbations must
have been part  of the initial conditions.\footnote{Note though that it
is not yet known for  definite that there are large-angle perturbations
intrinsic to the last  scattering surface. For example, in a topological
defect model such as  cosmic strings, such perturbations could be
generated as the microwave  photons propagate towards us.}

If this is the case, then the hot big bang theory does not allow a 
predictive theory for the origin of structure. While there is no reason
why  it is required to give a predictive theory, this would be a major
setback   and disappointment for the study of structure formation in the
Universe.

\subsection{The monopole problem (and other relics)}

Modern particle theories predict a variety of `unwanted relics', which would 
violate observations. These include
\begin{itemize}
\item Magnetic monopoles.
\item Domain walls.
\item Supersymmetric particles such as the gravitino.
\item `Moduli' fields associated with superstrings.
\end{itemize}
Typically, the problem is that these are expected to be created very early 
in the Universe's history, during the radiation era. But because they are 
diluted by the expansion more slowly than radiation (eg as
$a^{-3}$ instead of 
$a^{-4}$) it is very easy for them to become the dominant material in the 
Universe, in contradiction to observations. One has to dispose of them 
without harming the conventional matter in the Universe.

\section{The Idea of Inflation}

Seen with many years of hindsight, the idea of inflation is actually rather 
obvious. Take for example the Friedmann equation as used to analyze the 
flatness problem
\begin{equation}
|\Omega -1 | = \frac{|k|}{a^2 H^2} \,.
\end{equation}
The problem with the hot big bang model is that $aH$ always decreases, and 
so $\Omega$ is repelled away from one.

In order to solve the problem, we will clearly need to reverse this
state of affairs. Accordingly, 
define inflation to be any epoch where $\ddot{a} > 0$, an 
accelerated expansion. We can rewrite this in several different ways
\begin{eqnarray}
\mbox{INFLATION} \quad & \Longleftrightarrow \quad & \ddot{a} > 0 \\
  & \Longleftrightarrow \quad & \frac{d(H^{-1}/a)}{dt} < 0 \\
  & \Longleftrightarrow \quad & p < - \frac{\rho}{3}
\end{eqnarray}
The middle definition is the one which I prefer to use, 
because it has the most direct 
geometrical interpretation. It says that the Hubble length, as measured in 
comoving coordinates, {\em decreases} during inflation. At any other time, 
the comoving Hubble length increases. This is the key property of inflation; 
although typically the expansion of the Universe is very rapid, the crucial 
characteristic scale of the Universe is actually becoming smaller, when 
measured relative to that expansion.

As we will see, quite a wide range of behaviours satisfy the
inflationary condition. The most classic one is one we have already
seen; when the equation of state is $p=-\rho$, the solution is 
\begin{equation}
a(t) \propto \exp \left( Ht \right) \,.
\end{equation}

Since the successes of the hot big bang theory rely on the Universe
having a  conventional (non-inflationary) evolution, we cannot permit
this  inflationary period to go on forever --- it must come to an end
early enough  that the big bang successes are not threatened. Normally,
then, inflation is  viewed as a phenomenon of the very early Universe,
which comes to an end and  is followed by the conventional behaviour.
Inflation does not replace the  hot big bang theory; it is a bolt-on
accessory attached at early times to  improve the performance of the
theory.

\subsection{The flatness problem}

\begin{figure}[t]
\centering 
\leavevmode\epsfysize=6.3cm \epsfbox{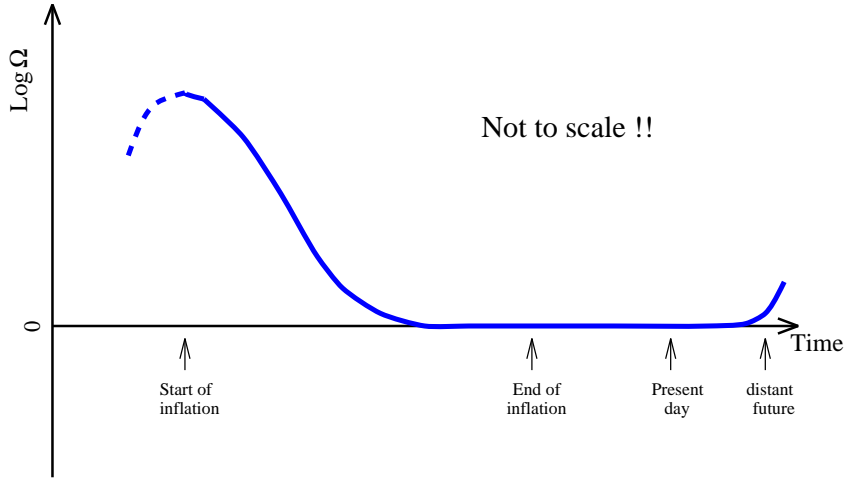}\\ 
\caption[flatness]{\label{arl:flatness} A possible evolution of $\Omega$. 
There 
may or may not be evolution before inflation, shown by the dotted line. 
During inflation $\Omega$ is forced dramatically towards one, and remains 
there right up to the present. Only in the extremely distant future will it 
begin to evolve away from one again.} 
\end{figure}

Inflation solves the flatness problem more or less by definition (so
that at least any {\em classical}, as opposed to quantum, solution of
the problem will fall under the umbrella of the inflationary
definition). From the middle condition, inflation is precisely the
 condition that $\Omega$ is forced towards one rather than away from it.
As  we shall see, this typically happens very rapidly. A short period of
such behaviour won't do us any good, as the subsequent non-inflationary
behaviour (in particular the standard big bang evolution from
nucleosynthesis onwards) will take us away from flatness again, but all
will be well provided we have  enough inflation that $\Omega$ is moved
extremely close to one during the  inflationary epoch. If it is close
enough, then it will stay very close to one right to the present, 
despite being repelled from one for all the post-inflationary period. 
Obtaining sufficient inflation to perform this task is actually fairly
easy. A schematic illustration of this behaviour is shown in
Figure~\ref{arl:flatness}.

In the above discussion, I have ignored a possible cosmological constant
contribution, but if present it modifies the Friedmann equation to
\begin{equation}
\left| \Omega + \Omega_\Lambda - 1 \right| = \frac{|k|}{a^2 H^2} \,,
\end{equation}
and so it is $\Omega + \Omega_\Lambda$ which is forced to one. In
general, it is spatial flatness ($k \simeq 0$) that we are driven
towards, not a critical matter density.

\subsection{Relic abundances}

The rapid expansion of the inflationary stage rapidly dilutes the
unwanted  relic particles, because the energy density during inflation
falls off more  slowly (as $a^{-2}$ or slower) than the relic particle
density. Very quickly  their density becomes negligible.

This resolution can only work if, after inflation, the energy density of
the  Universe can be turned into conventional matter without recreating
the  unwanted relics. This can be achieved by ensuring that during the 
conversion, known as {\em reheating}, the temperature never gets hot
enough  again to allow their thermal recreation. Then reheating can
generate solely  the things which we want. Such successful reheating
allows us to get back  into the hot big bang Universe, recovering all
its later successes such as  nucleosynthesis and the microwave
background.

\subsection{The horizon problem and homogeneity}

The inflationary expansion also solves the horizon problem. The basic 
strategy is to ensure that
\begin{equation}
\int_{t_*}^{t_{{\rm dec}}} \frac{dt}{a(t)} \gg \int_{t_{{\rm dec}}}^{t_0}
	\frac{dt}{a(t)} \,,
\end{equation}
so that light can travel much further before decoupling than it can 
afterwards. This cannot be done with standard evolution, but can be achieved 
by inflation.

\begin{figure}[t]
\centering 
\leavevmode\epsfysize=9.5cm \epsfbox{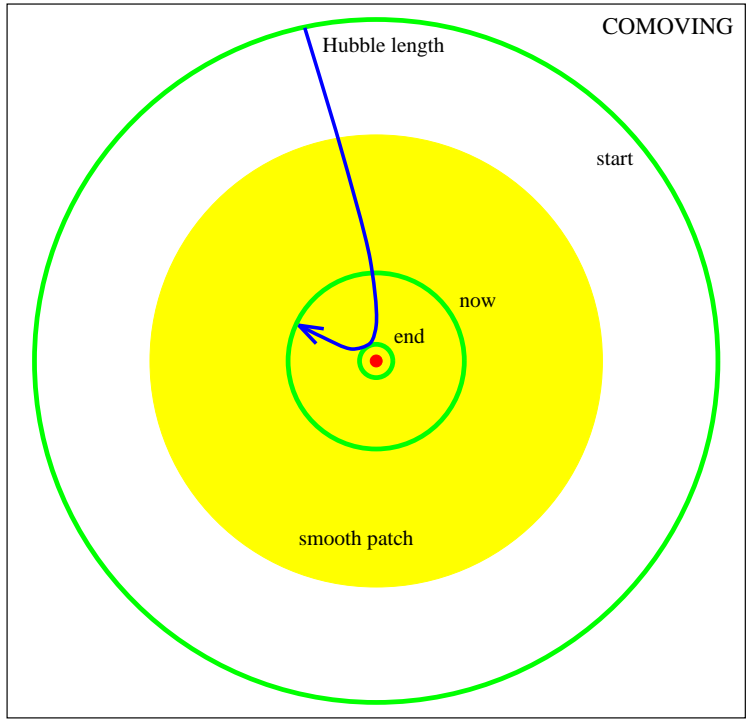}\\ 
\caption[horizon]{\label{arl:horizon} Solving the horizon problem. Initially 
the 
Hubble length is large, and a smooth patch forms by causal interactions. 
Inflation then shrinks the Hubble length, and even the subsequent expansion 
again after inflation leaves the observable Universe within the smoothed 
patch.} 
\end{figure} 

An alternative way to view this is to remember that inflation
corresponds to  a decreasing comoving Hubble length. The Hubble length
is ordinarily a good  measure of how far things can travel in the
Universe; what this is telling  us is that the region of the Universe we
can see after (even long after)  inflation is much smaller than the
region which would have been visible  before inflation started. Hence
causal physics was perfectly capable of  producing a large smooth
thermalized region, encompassing a volume greatly  in excess of our
presently observable Universe. In Figure~\ref{arl:horizon},  the  outer
circle indicates the initial Hubble length, encompassing the shaded 
smooth patch. Inflation shrinks this dramatically inwards towards the
dot  indicating our position, and then after inflation it increases
while staying  within the initial smooth patch.\footnote{Although this
is a standard  description, it isn't totally accurate. A more accurate
argument is as  follows.\cite{LLrep} At the beginning of inflation
particles are  distributed in a set of modes. This may be a thermal
distribution or  something else; whatever, since the energy density is
finite there will be a  shortest wavelength occupied mode, e.g. for a
thermal distribution  $\lambda_{{\rm max}} \sim 1/T$. Expressed in
physical coordinates, once  inflation has stretched all modes including
this one to be much larger than  the Hubble length, the Universe becomes
homogeneous. In comoving  coordinates, the equivalent picture is that
the Hubble length shrinks in  until it's much smaller than the shortest
wavelength, and the Universe, as  before, appears homogeneous.}

Equally, causal processes would be capable of generating irregularities
in  the Universe on scales greatly exceeding our presently observable
Universe, provided they happened at an early enough time that those
scales were within causal contact. This will be explored in detail
later.

\section{Modelling the Inflationary Expansion}

We have seen that a period of accelerated expansion --- inflation --- is
sufficient to resolve a range of cosmological problems. But we need a 
plausible scenario for driving such an expansion if we are to be able to
make proper calculations. This is provided by cosmological scalar
fields.

\subsection{Scalar fields and their potentials}

In particle physics, a scalar field is used to represent spin zero 
particles. It transforms as a scalar (that is, it is unchanged) under 
coordinate transformations. In a homogeneous Universe, the scalar field
is a  function of time alone.

In particle theories, scalar fields are a crucial ingredient for
spontaneous  symmetry breaking. The most famous example is the Higgs
field which breaks  the electro-weak symmetry, whose existence is hoped
to be verified at the  Large Hadron Collider at CERN when it commences
experiments next millennium.  Scalar fields are also expected to be
associated with the breaking of other  symmetries, such as those of
Grand Unified Theories, supersymmetry etc.
\begin{itemize}
\item Any specific particle theory (eg GUTS, superstrings) contains scalar 
fields.
\item No fundamental scalar field has yet been observed.
\item In condensed matter systems (such as superconductors, superfluid 
helium etc) scalar fields are widely observed, associated with any phase 
transition. People working in that subject normally refer to the scalar 
fields as `order parameters'.
\end{itemize}

The traditional starting point for particle physics models is the
action, which is an integral of the Lagrange density over space and time
and from which the equations of motion can be obtained. As an
intermediate step, one might write down the energy--momentum tensor,
which sits on the right-hand side of Einstein's equations. Rather than
begin there, I will take as my starting point expressions for
the  effective energy density and pressure of a homogeneous scalar
field, which  I'll call $\phi$. These are obtained by comparison of the
energy--momentum tensor of the scalar field with that of a perfect
fluid, and are
\begin{eqnarray}
\label{arl:effrho}
\rho_{\phi} & = & \frac{1}{2} \dot{\phi}^2 + V(\phi) \, \\
\label{arl:effp}
p_{\phi} & = & \frac{1}{2} \dot{\phi}^2 - V(\phi) \,.
\end{eqnarray}
One can think of the first term in each as a kinetic energy, and the
second  as a potential energy. The potential energy $V(\phi)$ can be
thought of as a  form of `configurational' or `binding' energy; it
measures how much internal  energy is associated with a particular field
value. Normally, like all  systems, scalar fields try to minimize this
energy; however, a crucial  ingredient which allows inflation is that
scalar fields are not always very  efficient at reaching this minimum
energy state.

Note in passing that a scalar field cannot in general be described by an 
equation of state; there is no unique value of $p$ that can be associated 
with a given $\rho$ as the energy density can be divided between potential 
and kinetic energy in different ways.

\begin{figure}[t]
\centering 
\leavevmode\epsfysize=5.5cm \epsfbox{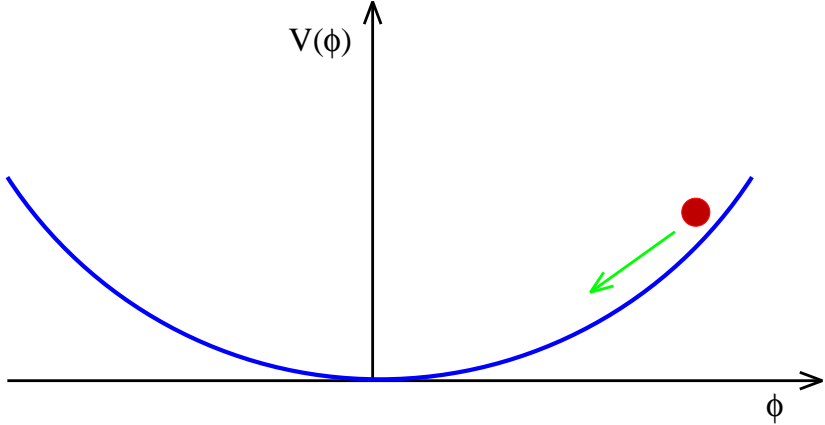}\\ 
\caption[scalpot]{\label{arl:scalpot} A generic inflationary potential.} 
\end{figure}

In a given theory, there would be a specific form for the potential 
$V(\phi)$, at least up to some parameters which one could hope to measure 
(such as the effective mass and interaction strength of the scalar field). 
However, we are not presently in a position where there is a well 
established fundamental theory that one can use, so, in the absence of such 
a theory, inflation workers tend to regard $V(\phi)$ as a function to be 
chosen arbitrarily, with different choices corresponding to different models 
of inflation (of which there are many). Some example potentials are
\begin{eqnarray}
& V(\phi) = \lambda \left( \phi^2 - M^2 \right)^2 \quad \quad & 
	\mbox{Higgs potential} \\
& V(\phi) = \frac{1}{2} m^2 \phi^2 & \mbox{Massive scalar field}\\
& V(\phi) = \lambda \phi^4 & \mbox{Self-interacting scalar field}
\end{eqnarray}
The strength of this approach is that it seems possible to capture many of 
the crucial properties of inflation by looking at some simple potentials; 
one is looking for results which will still hold when more `realistic' 
potentials are chosen. Figure~\ref{arl:scalpot} shows such a generic 
potential, 
with the scalar field displaced from the minimum and trying to reach it.

\subsection{Equations of motion and solutions}

The equations for an expanding Universe containing a homogeneous scalar 
field are easily obtained by substituting Eqs.~(\ref{arl:effrho}) and 
(\ref{arl:effp}) into the Friedmann and fluid equations, giving
\begin{eqnarray}
\label{arl:scalfried}
H^2 & = & \frac{8\pi}{3 \mpl^2} \left[ V(\phi) + \frac{1}{2} \dot{\phi}^2
	\right] \,, \\
\label{arl:scalwave}
\ddot{\phi} + 3 H \dot{\phi} & = & - V'(\phi) \,,
\end{eqnarray}
where prime indicates $d/d\phi$. Here I have ignored the curvature term
$k$, since we know that by definition it will quickly become negligible
once inflation starts. This is done for simplicity only; there is no
obstacle to including that term.

Since
\begin{equation}
\ddot{a} > 0 \Longleftrightarrow p < - \frac{\rho}{3} \Longleftrightarrow 
       \dot{\phi}^2 < V(\phi)
\end{equation}
we will have inflation whenever the potential energy dominates. This should 
be possible provided the potential is flat enough, as the scalar field would 
then be expected to roll slowly. The potential should also have a minimum in 
which inflation can end.

The standard strategy for solving these equations is the {\bf slow-roll 
approximation} (SRA); this assumes that a term can be neglected in each of 
the equations of motion to leave the simpler set
\begin{eqnarray}
H^2 & \simeq & \frac{8\pi}{3 \mpl^2} \, V \\
3 H \dot{\phi} & \simeq & -V'
\end{eqnarray}
If we define {\bf slow-roll parameters}~\cite{LL}
\begin{equation}
\label{arl:SR}
\epsilon(\phi) = \frac{\mpl^2}{16 \pi} \, \left( \frac{V'}{V}
	\right)^2 \quad ; \quad \eta(\phi) = \frac{\mpl^2}{8\pi}
	\, \frac{V''}{V} \,,
\end{equation}
where the first measures the slope of the potential and the second the 
curvature, then necessary conditions for the slow-roll approximation to hold 
are~\footnote{Note that $\epsilon$ is positive by definition, whilst $\eta$ 
can have either sign.}
\begin{equation}
\epsilon \ll 1 \quad ; \quad |\eta| \ll 1 \,.
\end{equation}
Unfortunately, although these are necessary conditions for the slow-roll 
approximation to hold, they are not sufficient, since even if the
potential is very flat it may be that the scalar field has a large
velocity. A more elaborate version of the SRA exists, based on the
Hamilton--Jacobi formulation of inflation,\cite{SB} which is sufficient
as well as necessary.\cite{LPB}

Note also that the SRA reduces the order of the system of equations by
one, and so  its general solution contains one less initial condition.
It works only  because one can prove~\cite{SB,LPB} that the solution to
the full equations  possesses an attractor property, eliminating the
dependence on the extra  parameter.

\subsection{The relation between inflation and slow-roll}

\label{arl:infsr}

As it happens, the applicability of the slow-roll condition is closely
connected to the condition for inflation to take place, and in many
contexts the conditions can be regarded as equivalent. Let's quickly see
why. 

The inflationary condition $\ddot{a} > 0$ is satisfied for a much wider 
range of behaviours than just (quasi-)exponential expansion. A classic 
example is power-law inflation $a \propto t^p$ for $p>1$, which is an exact 
solution for an exponential potential
\begin{equation}
V(\phi) = V_0 \exp \left[ - \sqrt{\frac{16 \pi}{p}} \, \frac{\phi}{\mpl}
	\right] \,.
\end{equation}

We can manipulate the condition for inflation as
\begin{eqnarray}
& &\frac{\ddot{a}}{a} = \dot{H} + H^2 > 0  \nonumber \\
\Longleftrightarrow & & -\frac{\dot{H}}{H^2} < 1 \nonumber \\
\stackrel{\sim}{\Longleftrightarrow} & & \frac{\mpl^2}{16\pi} \, \left( 
	\frac{V'}{V} \right)^2 < 1 \nonumber 
\end{eqnarray}
where the last manipulation uses the slow-roll approximation. The final 
condition is just the slow-roll condition $\epsilon < 1$, and hence
\[
\mbox{Slow-roll} \Longrightarrow \mbox{Inflation}
\]
Inflation will occur when the slow-roll conditions are satisfied (subject
to some caveats on whether the `attractor' behaviour has been
attained.\cite{LPB}) 

However, the converse is not strictly true, since we had to use the 
SRA in the derivation. However, in practice
\begin{eqnarray}
\mbox{Inflation} & \stackrel{\sim}{\Longrightarrow} & \epsilon < 1 
	\nonumber \\
\mbox{Prolonged inflation} & \stackrel{\sim}{\Longrightarrow} & \eta < 1 
	\nonumber
\end{eqnarray}
The last condition arises because unless the curvature of the potential
is small, the potential will not be flat for a wide enough range of
$\phi$. 

\subsection{The amount of inflation}

The amount of inflation is normally specified by the logarithm of the amount 
of expansion, {\em the number of e-foldings} $N$, given by
\begin{eqnarray}
N \equiv \ln \frac{a(t_{{\rm end}})}{a(t_{{\rm initial}})} & = &
	\int_{t_{{\rm i}}}^{t_{{\rm e}}} \, H \, dt \,, \\
 & \simeq & - \frac{8\pi}{\mpl^2} \int_{\phi_{{\rm i}}}^{\phi_{{\rm e}}}
 	\, \frac{V}{V'} \, d\phi \,, \label{efold}
\end{eqnarray}
where the final step uses the SRA. Notice that the amount of inflation 
between two scalar field values can be calculated without needing to solve 
the equations of motion, and also that it is unchanged if one multiplies 
$V(\phi)$ by a constant.

The minimum amount of inflation required to solve the various
cosmological  problems is about 70 $e$-foldings, i.e.~an expansion by a
factor of  $10^{30}$. Although this looks large, inflation is typically
so rapid that  most inflation models give much more.

\subsection{A worked example: polynomial chaotic inflation}

\label{arl:quadratic}

The simplest inflation model~\cite{Linbook} arises when one chooses a 
polynomial potential, such as that for a massive but otherwise 
non-interacting field, $V(\phi) = m^2 \phi^2/2$ where $m$ is the mass of the 
scalar field. With this potential, the slow-roll equations are
\begin{equation}
3H\dot{\phi} + m^2 \phi = 0 \quad ; \quad H^2 = \frac{4\pi m^2
	\phi^2}{3\mpl^2} \,,
\end{equation}
and the slow-roll parameters are
\begin{equation}
\epsilon = \eta = \frac{\mpl^2}{4 \pi \phi^2} \,.
\end{equation}
So inflation can proceed provided $|\phi| > \mpl/\sqrt{4\pi}$, i.e.~as
long as we are not to close to the minimum. 

The slow-roll equations are readily solved to give
\begin{eqnarray}
\phi(t) & = & \phi_{{\rm i}} - \frac{m \, \mpl}{\sqrt{12\pi}} \, t \,, \\
a(t) & = & a_{{\rm i}} \exp \left[ \sqrt{\frac{4\pi}{3}} \, \frac{m}{\mpl}
	\, \left( \phi_{{\rm i}} t - \frac{m \, \mpl}{\sqrt{48\pi}} t^2
	\right) \right] \,,
\end{eqnarray}
(where $\phi = \phi_{{\rm i}}$ and $a = a_{{\rm i}}$ at $t = 0$) and the 
total amount of inflation is
\begin{equation}
\label{arl:quadefold}
N_{{\rm tot}} = 2 \pi \, \frac{\phi_{{\rm i}}^2}{\mpl^2} - \frac{1}{2} \,.
\end{equation}
This last equation can be obtained from the solution for $a$, but in
fact is more easily obtained directly by integrating Eq.~(\ref{efold}),
for which one needn't bother to solve the equations of motion.

In order for classical physics to be valid we require $V \ll \mpl^4$,
but it  is still easy to get enough inflation provided $m$ is small
enough. As we shall later see, $m$ is in fact required to be small from
observational limits on the size of density perturbations produced, and
we can easily get far more than the minimum amount of inflation required
to solve the various cosmological problems we originally set out to
solve. 

\subsection{Reheating after inflation}

During inflation, all matter except the scalar field (usually called the
 inflaton) is redshifted to extremely low densities. {\bf Reheating} is
the  process whereby the inflaton's energy density is converted back
into conventional matter after inflation, re-entering the standard big
bang theory.

Once the slow-roll conditions break down, the scalar field switches from
being overdamped to being underdamped and begins to move  rapidly on the
Hubble timescale, oscillating at the bottom of the  potential. As it
does so, it decays into conventional matter. The details of  reheating
are an important area of research in inflationary cosmology at the
moment for several reasons, but are not important for the generation and
evolution of density perturbations which is the main focus of the
remainder of this article. Consequently, I'll just note that recently
there has been quite a dramatic change of view as to how reheating takes
place. Traditional treatments (e.g.~as given in Kolb \&
Turner~\cite{KT}) added a phenomenological decay term; this  was
constrained to be very small and hence reheating was viewed as being 
very inefficient. This allowed substantial redshifting to take place
after  the end of inflation and before the Universe returned to thermal 
equilibrium; hence the reheat temperature would be lower, by several
orders  of magnitude, than suggested by the energy density at the end of
inflation.

This picture is radically revised in work by Kofman, Linde \&
Starobinsky~\cite{KLS} (see also Ref.~\cite{preheat}), who suggest that
the decay can undergo broad parametric resonance, with extremely
efficient transfer of energy from the coherent oscillations of the
inflaton field. This initial  transfer has been dubbed {\em preheating}.
With such an efficient start to  the reheating process, it now appears
possible that the reheating epoch may  be very short indeed and hence
that most of the energy density in the  inflaton field at the end of
inflation may be available for conversion into  thermalized form. 

\subsection{The range of inflation models}

Over the last fifteen years or so a great number of inflationary models
have  been devised, both with and without reference to specific
underlying particle theories. Here I will discuss a very small subset
of the models which have been introduced, just to give you a flavour of
the variety. At the moment particle physics model building of inflation
is undergoing a renaissance, and a detailed snapshot of the current
situation can be found in the review of Lyth \& Riotto.\cite{LR}

However, as we shall be discussing in the next section, observations
have  great prospects for distinguishing between the different
inflationary  models. By far the best type of observation for this
purpose appears to be  high resolution satellite microwave background
anisotropy observations, and  we are fortunate that two proposals have
been approved --- NASA has  funded the {\em MAP} satellite~\cite{MAP} 
for launch around 2000, and ESA  has approved the {\sc
Planck} satellite~\cite{planck} for launch some later. These
satellites should offer  very strong discrimination between the
inflation models I shall now discuss.  Indeed, it may even be possible
to attempt a more challenging type of  observation --- one which is
independent of the particular inflationary  model and hence begins to
test the idea of inflation itself.

\subsubsection{Chaotic inflation models}

This is the standard type of inflation model.\cite{Linbook} The ingredients 
are
\begin{itemize}
\item A single scalar field, rolling in ...
\item A potential $V(\phi)$, which in some regions satisfies the slow-roll 
conditions, while also possessing a minimum with zero potential in which 
inflation is to end.
\item Initial conditions well up the potential, due to large fluctuations at 
the Planck era.
\end{itemize}
There are a large number of models of this type. Some are
\begin{quote}
\begin{tabbing}
Polynomial chaotic inflation \hspace*{0.3cm} \= $V(\phi) = 
	\frac{1}{2} m^2 \phi^2$ \\
 \> $V(\phi) = \lambda \phi^4$ \\
Power-law inflation \> $V(\phi) = V_0 \exp (\sqrt{\frac{16\pi}{p}} \,
	\frac{\phi}{\mpl})$ \\
`Natural' inflation \> $V(\phi) = V_0 [ 1 + \cos \frac{\phi}{f} ]$ \\
Intermediate inflation \> $V(\phi) \propto \phi^{-\beta}$ 
\end{tabbing}
\end{quote}
Some of these actually do not satisfy the condition of a minimum in which 
inflation ends; they permit inflation to continue forever. However, we shall 
see power-law inflation arising in a more satisfactory context shortly.

\subsubsection{Multi-field theories}

\begin{figure}[tb]
\centering 
\leavevmode\epsfysize=6.5cm \epsfbox{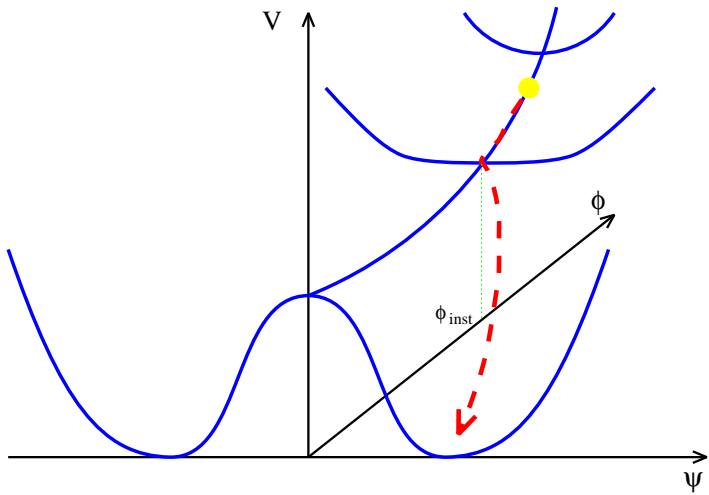}\\ 
\caption[hybridpot]{\label{arl:hybridpot} The potential for the hybrid 
inflation model. The field rolls down the channel at $\psi = 0$ until it 
reaches the critical $\phi$ value, then falls off the side to the true 
minimum at $\phi = 0$ and $\psi = \pm M$.} 
\end{figure}

A recent trend in inflationary model building has been the exploration of 
models with more than one scalar field. The classic example is the hybrid 
inflation model,\cite{hybrid} which seems particularly promising for 
particle physics model building. The simplest version has a potential 
with two fields $\phi$ and $\psi$ of the form
\begin{equation}
V(\phi,\psi) = \frac{\lambda}{4} \left( \psi^2 - M^2 \right)^2 + 
	\frac{1}{2} m^2 \phi^2 + \frac{1}{2} \lambda' \phi^2 \psi^2 \,.
\end{equation}
which is illustrated in Figure~\ref{arl:hybridpot}. When $\phi^2$ is large, 
the 
minimum of the potential in the $\psi$-direction
is at $\psi = 0$. The field rolls down this 
`channel' until it reaches $\phi_{{\rm inst}}^2 = \lambda M^2/\lambda'$, at 
which point $\psi = 0$ becomes unstable and the field rolls into one of the 
true minima at $\phi = 0$ and $\psi = \pm M$.

While in the `channel', which is where all the interesting behaviour takes 
place, this is just like a single field model with an effective potential 
for $\phi$ of the form
\begin{equation}
V_{{\rm eff}}(\phi) = \frac{\lambda}{4} M^4 + \frac{1}{2} m^2 \phi^2 \,.
\end{equation}
This is a fairly standard form, the unusual thing being the constant term, 
which would not normally be allowed as it would give a present-day 
cosmological constant. The most interesting regime is where that constant 
dominates, and it gives quite an unusual phenomenology. In particular, the 
energy density during inflation can be much lower than normal while still 
giving suitably large density perturbations, and secondly the field $\phi$ 
can be rolling extremely slowly which is of benefit to particle physics 
model building.

Within the more general class of two and multi-field inflation models,
it is  quite common for only one field to be dynamically important, as
in the  hybrid inflation model --- this effectively reduces the
situation back to  the single field case of the previous subsection.
However, it may also be  possible to have more than one important
dynamical degree of freedom. In  that case there is no attractor
behaviour giving a unique route into the  potential minimum, as in the
single field case; for example, if the  potential is of the form of an
asymmetric bowl one could roll into the base  down any direction. In
that situation, the model loses some of its predictive power,  because
the late-time behaviour is not independent of the initial 
conditions.\footnote{Of course, there is no requirement that the `true' 
physical theory does have predictive power, but it would be unfortunate
for  us if it does not.}

\subsubsection{Beyond general relativity}

Rather than introduce an explicit scalar field to drive inflation, some 
theories modify the gravitational sector of the theory into something more 
complicated than general relativity.\cite{W93} Examples are
\begin{itemize}
\item Higher derivative gravity  ($R + R^2 + \cdots$).
\item Jordan--Brans--Dicke theory.
\item Scalar--tensor gravity.
\end{itemize}
The last two are theories where the gravitational constant may vary (indeed 
Jordan--Brans--Dicke theory is a special case of scalar--tensor gravity).

However, a clever trick, known as the {\em conformal
transformation},\cite{conform} allows such theories to be rewritten as
general relativity  plus one or more scalar fields with some potential.
Often, only one of those  fields is dynamical which returns us once more
to the original chaotic  inflation scenario!

The most famous example is extended inflation.\cite{extinf} In its
original  form, it transforms precisely into the power-law inflation
model that we've  already discussed, with the added bonus that it
includes a proper method of  ending inflation. Unfortunately though,
this model is now ruled out by  observations.\cite{LL} Indeed, models
of inflation based on altering  gravity are much more constrained than
other types, since we know a lot  about gravity and how well general
relativity works,\cite{W93} and many  models of this kind are very
vulnerable to observations.

\subsubsection{Open inflation}

In the early 1990s, in the face of ever increasing evidence of a
sub-critical matter density in the Universe, interest was refocussed on
an idea which defies the original inflationary motivation and gives rise
to a homogeneous but open Universe from inflation.\footnote{That is, a
genuinely open Universe with hyperbolic geometry and no cosmological
constant.} Often in the past it has been declared that this is either
impossible or contrived; however, it can be readily achieved in models
with quantum tunnelling from a false vacuum (a metastable state)
followed by a second inflationary stage.\cite{open} The tunnelling
creates a bubble, and, incredibly, the region inside the expanding
bubble looks just like an open Universe, with the bubble wall 
corresponding to the initial (coordinate) singularity. These models are 
normally referred to as `open inflation' or `single-bubble' models. So
far it has turned out that such models are not all that easy to
construct. 

These models are already very different from traditional inflation
models, and subsequently an even bolder idea has been proposed,\cite{HT}
that an open Universe can be created via `tunnelling from nothing'
rather than from a pre-existing inflationary phase. As I write this
remains controversial.

While both these types of open inflation models remain viable, they are
considerably more complex than the standard inflation models, and at the
moment not that well motivated as although observations continue to
favour a low matter density, they also favour spatial flatness
reintroduced by a cosmological constant. Therefore from now on I will
restrict discussion  to the single-field chaotic inflation models.

\subsection{Recap}

The main points of this long section were the following.
\begin{itemize}
\item Cosmological scalar fields, which were introduced long before 
inflation was thought of, provide a natural framework for inflation.
\item Despite a wide range of motivations, most inflationary models are 
dynamically equivalent to general relativity plus a single scalar field with 
some potential $V(\phi)$.
\item Within this framework, solutions describing inflation are easily 
found. Indeed, for many of the properties (amount of expansion, for 
example), we do not even need to solve the equations of motion.
\end{itemize}
With this information under our belts, we are now able to discuss the 
strongest motivation for the inflationary cosmology --- that it is able to 
provide an explanation for the origin of structure in the Universe.

\section{Density Perturbations and Gravitational Waves}

In modern terms, by far the most important property of inflationary 
cosmology is that it produces spectra of both density perturbations and 
gravitational waves. The density perturbations may be responsible for the 
formation and clustering of galaxies, as well as creating anisotropies in 
the microwave background radiation. The gravitational waves do not affect 
the formation of galaxies, but as we shall see may contribute extra 
microwave anisotropies on the large angular scales sampled by the COBE 
satellite.\cite{COBE1,COBE4} An alternative terminology for the density
perturbations is scalar perturbations and for the gravitational waves is
tensor perturbations, the terminology referring to their transformation
properties. 

Studies of large-scale structure typically make some assumption about the 
initial form of these spectra. Usually gravitational waves are assumed not 
to be present, and the density perturbations to take on a simple form such 
as the scale-invariant Harrison--Zel'dovich spectrum, or a scale-free 
power-law spectrum. It is clearly highly desirable to have a theory which 
predicts the forms of the spectra. There are presently two rival models 
which do this, {\em cosmological inflation} and {\em topological defects}.
At present inflation is favoured both on observational grounds and
because it provides a simpler framework for understanding the evolution
of structure

\subsection{Production during inflation}

The ability of inflation to generate perturbations on large scales comes
from the unusual behaviour of the Hubble length during inflation, namely
that (by definition) the comoving Hubble length decreases. When we talk
about large-scale structure, we are primarily interested in comoving
scales,  as to a first approximation everything is dragged along with
the expansion.  The qualitative behaviour of irregularities is governed
by their scale in  comparison to the characteristic scale of the
Universe, the Hubble length. 

In the big bang Universe the comoving Hubble length is always
increasing,  and so all scales are initially much larger than it, and
hence unable to be  affected by causal physics. Once they become smaller
than the Hubble length,  they remain so for all time. In the standard
scenarios, COBE sees  perturbations on large scales at a time when they
were much bigger than the  Hubble length, and hence no mechanism could
have created them.

\begin{figure}[t!]
\centering 
\leavevmode\epsfysize=12cm \epsfbox{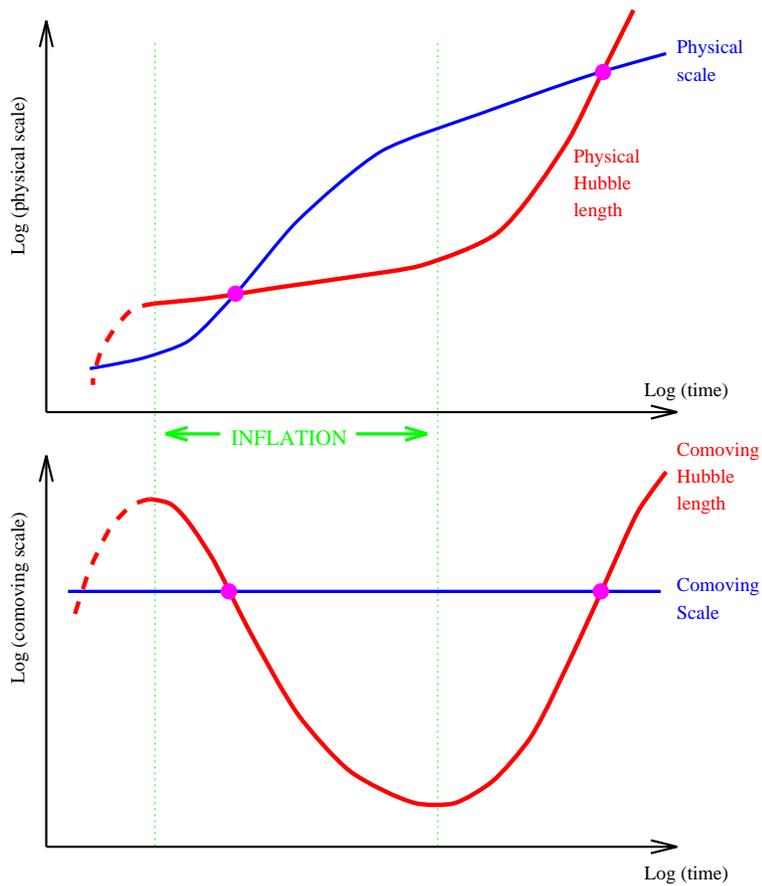}\\ 
\caption[scales]{\label{arl:scales} The behaviour of a given comoving scale 
relative to the Hubble length, both during and after inflation, shown 
using physical coordinates (upper panel) and comoving ones (lower panel).} 
\end{figure} 

Inflation reverses this behaviour, as seen in Figure~\ref{arl:scales}.
Now a given comoving scale has a more complicated history. Early on in
inflation, the scale could be well inside the Hubble length, and hence
causal physics  can act, both to generate homogeneity to solve the
horizon problem and to  superimpose small perturbations. Some time
before inflation ends, the scale  crosses outside the Hubble radius
(indicated by a circle in the lower panel  of Figure~\ref{arl:scales})
and causal physics becomes ineffective. Any  perturbations generated
become imprinted, or, in the usual  terminology, `frozen in'. Long after
inflation is over, the scales cross  inside the Hubble radius again.
Perturbations are created on a very wide range of scales, but the most
readily observed ones range  from about the size of the
present Hubble radius (i.e.~the size of the  presently observable
Universe) down to a few orders of magnitude less. On  the scale of
Figure~\ref{arl:scales}, all interesting comoving scales lie  extremely
close together, and cross the Hubble radius during inflation very  close
together.

It's all very well to realize that the dynamics of inflation permits 
perturbations to be generated without violating causality, but we need a
 specific mechanism. That mechanism is quantum fluctuations. Inflation
is  trying as hard as it can to make the Universe perfectly homogeneous,
but it  cannot defeat the Uncertainty Principle which ensures that there
are always  some irregularities left over. Through this limitation, it
is possible for  inflation to adequately solve the homogeneity problem
and in addition  leave enough irregularities behind to attempt to
explain why the present  Universe is not completely homogeneous. 

The size of the irregularities depends on the energy scale at which 
inflation takes place. It is outside the scope of these lectures to describe 
in detail how this calculation is performed (see e.g.~Ref.~\cite{LLKCBA} for 
a reasonably accessible description); 
I'll just briefly outline the necessary steps and then 
quote the result, which we can go on to apply.
\begin{quote}
\begin{tabbing}
(a)~Perturb the scalar field \hspace*{3.5cm} \= $\phi = \phi(t) +
	\delta \! \phi(\x,t)$ \\
(b)~Expand in comoving wavenumbers \> $\delta \phi = \sum
	(\delta \! \phi)_{\k} e^{i\k.\x}$ \\
(c)~Linearized equation for classical evolution \> 
	 \\
(d)~Quantize theory   \\
(e)~Find solution with initial condition giving  \\
~~~~~flat space quantum theory ($k \gg aH$)  \\
(f)~Find asymptotic value for $k \ll aH$ \> $\langle | \delta \!
	\phi_{\k}|^2 \rangle = H^2/2k^3$ \\
(g)~Relate field perturbation to metric \> ${\cal R} = H \, \delta
	\! \phi/\dot{\phi}$ \\
~~~~~or curvature perturbation
\end{tabbing}
\end{quote}
Some important points are
\begin{itemize}
\item The details of this calculation are extremely similar to those used to 
calculate the Casimir effect (a quantum force between parallel plates), 
which has been tested in the laboratory.
\item The calculation itself is not controversial, though some aspects of 
its interpretation (in particular concerning the quantum to classical 
transition) are.
\item Exact analytic results are not known for general inflation models 
(though linear theory results for arbitrary models are readily calculated 
numer\-ically~\cite{GL}). 
The results I'll be quoting will be lowest-order in the SRA, 
which is good enough for present observations. 
\item Results are known to second-order in slow-roll for arbitrary inflaton 
potentials.\cite{SL} Power-law inflation is the only standard model for which 
exact 
results are known. In some other cases, high accuracy approximations give 
better results (e.g.~small-angle approximation in natural or hybrid 
inflation~\cite{SL,GW}).
\end{itemize}

The formulae for the amplitude of density perturbations, which I'll call 
$\delh(k)$, and the gravitational waves, $A_{{\rm G}}(k)$, are~\footnote{The 
precise normalization of the spectra is arbitrary, as are the number of 
powers of $k$ included. I've made my favourite choice here (following 
Refs.~\cite{LLrep,LLKCBA}), 
but whatever convention is used the normalization 
factor will disappear in any physical answer. For reference, the usual 
power spectrum $P(k)$ is proportional to $k \delh^2(k)$.}
\begin{eqnarray}
\label{arl:delh}
\delh(k) & = & \left. \sqrt{\frac{512 \pi}{75}} \, \frac{V^{3/2}}{\mpl^3
	|V'|} \right|_{k = aH} \,, \\
\label{arl:AG}
A_{{\rm G}}(k) & = & \left. \sqrt{\frac{32}{75}} \, \frac{V^{1/2}}{\mpl^2}
	\right|_{k = aH} \,.
\end{eqnarray}
Here $k$ is the comoving wavenumber; the perturbations are normally analyzed 
via a Fourier expansion into comoving modes. The right-hand sides of the 
above equations are to be evaluated at the time when $k = aH$ 
during inflation, which for a given $k$ corresponds to some particular value 
of $\phi$. We see that the amplitude of perturbations depends on the 
properties of the inflaton potential at the time the scale crossed the 
Hubble radius during inflation. The relevant number of $e$-foldings from the 
end of inflation is given by~\cite{LLrep}
\begin{equation}
N \simeq 62 - \ln \frac{k}{a_0 H_0} + \mbox{numerical correction} \,,
\end{equation}
where `numerical correction' is a typically smallish (order a few) 
number which depends on the energy scale of inflation, the duration of 
reheating and so on. Normally it is a perfectly fine approximation to say 
that the scales of interest to us crossed outside the Hubble radius 60 
$e$-foldings before the end of inflation. Then the $e$-foldings formula
\begin{equation}
\label{arl:efold2}
N \simeq - \frac{8\pi}{\mpl^2} \int_{\phi}^{\phi_{{\rm end}}} \, 
	\frac{V}{V'} \, d\phi \,,
\end{equation}
tells us the value of $\phi$ to be substituted into Eqs.~(\ref{arl:delh}) 
and (\ref{arl:AG}).

\subsection{A worked example}

The easiest way to see what is going on is to work through a specific 
example, the $m^2 \phi^2/2$ potential which we already saw in 
Section~\ref{arl:quadratic}. We'll see that we don't even have to solve the 
evolution equations to get our predictions.
\begin{enumerate}
\item Inflation ends when $\epsilon = 1$, so $\phi_{{\rm end}} \simeq 
\mpl/\sqrt{4\pi}$.
\item We're interested in 60 $e$-foldings before this, which from 
Eq.~(\ref{arl:quadefold}) gives $\phi_{60} \simeq 3 \mpl$.
\item Substitute this in:
\[
\delh \simeq 12 \, \frac{m}{\mpl} \quad ; \quad A_{{\rm G}} \simeq 1.4
	\, \frac{m}{\mpl} 
\]
\item Reproducing the COBE result requires~\cite{BLW} 
$\delh \simeq 2 \times 10^{-5}$ 
(provided $A_{{\rm G}} \ll \delh$), so we need $m \simeq 10^{-6} 
\mpl$.
\end{enumerate}

Because the required value of $m$ is so small, that means it is easy to
get sufficient inflation to solve the cosmological problems, without
violating the classicality condition $V< \mpl^4$. That implies only that
$\phi < \mpl^2/m \simeq 10^6 \mpl$, and as $N_{{\rm tot}} \simeq 2\pi
\phi^2/\mpl^2$, we can get up to about $10^{13}$ $e$-foldings in
principle. This compares extremely favourably with the 70 or so actually
required. 

\subsection{Observational consequences}

Observations have moved on beyond us wanting to know the overall 
normalization of the potential. The interesting things are
\begin{enumerate}
\item The scale-dependence of the spectra.
\item The relative influence of the two spectra.
\end{enumerate}
These can be neatly summarized using the slow-roll parameters $\epsilon$
and  $\eta$ we defined earlier.\cite{LL}

The standard approximation used to describe the spectra is the {\bf 
power-law approximation}, where we take
\begin{equation}
\delh^2(k) \propto k^{n-1} \quad ; \quad A_{{\rm G}}^2(k) 
	\propto k^{n_{{\rm G}}} \,,
\end{equation}
where the spectral indices $n$ and $n_{{\rm G}}$ are given by
\begin{equation}
n-1 = \frac{d \ln \delh^2}{d \ln k} \quad ; \quad n_{{\rm G}} =
	\frac{d \ln A_{{\rm G}}^2}{d \ln k} \,.
\end{equation}
The power-law approximation is usually valid because only a limited
range of  scales are observable, with the range $1$ Mpc to $10^4$ Mpc
corresponding to  $\Delta \ln k \simeq 9$.

The crucial equation we need is that relating $\phi$ values to when a scale 
$k$ crosses the Hubble radius, which from Eq.~(\ref{arl:efold2}) is
\begin{equation}
\frac{d \ln k}{d \phi} = \frac{8\pi}{\mpl^2} \, \frac{V}{V'} \,.
\end{equation}
(since within the slow-roll approximation $k \simeq \exp N$). Direct 
differentiation then yields~\cite{LL}
\begin{eqnarray}
n & = & 1 - 6\epsilon + 2 \eta \,, \\
n_{{\rm G}} & = & -2 \epsilon \,,
\end{eqnarray}
where now $\epsilon$ and $\eta$ are to be evaluated on the appropriate part 
of the potential.

Finally, we need a measure of the relevant importance of density 
perturbations and gravitational waves. The natural place to look is the 
microwave background; a detailed calculation which I cannot reproduce here 
(see e.g.~Ref.~\cite{LLrep}) gives
\begin{equation}
\label{arl:relgrav}
R \equiv \frac{C_{\ell}^{{\rm GW}}}{C_{\ell}^{{\rm DP}}} \simeq \,
	4 \pi \epsilon \,.
\end{equation}
Here the $C_{\ell}$ are the contributions to the microwave multipoles, in 
the usual notation.\footnote{Namely, $\Delta T/T = \sum a_{\ell m} 
Y^{\ell}_{m}(\theta,\phi)$, $C_{\ell} = \langle | a_{\ell m}|^2 \rangle$.}

From these expressions we immediately see
\begin{itemize}
\item If and only if $\epsilon \ll 1$ and $|\eta| \ll 1$ do we get $n \simeq 
1$ and $R \simeq 0$.
\item Because the coefficient in Eq.~(\ref{arl:relgrav}) is so large, 
gravitational waves can have a significant effect even if $\epsilon$ is 
quite a bit smaller than one.
\end{itemize}

Table~\ref{arl:pred} shows the predictions for a range of inflation
models. The information I've given you so far should be sufficient to
allow you to reproduce them. Even  the simplest inflation models can
affect the large-scale structure modelling at a level comparable to the
present observational accuracy. The predictions of the different models
will be wildly different as far as future high-accuracy observations are
concerned.

\begin{table}
\begin{tabular}{|l|c|c|c|}
\hline
MODEL & POTENTIAL & $n$ & $R$\\
\hline
\hline
Polynomial        & $\phi^2$ & 0.97 & 0.1 \\
chaotic inflation & $\phi^4$ & 0.95 & 0.2 \\
\hline
Power-law inflation & $\exp ( -\lambda \phi)$ & any $n<1$ & $2\pi(1-n)$ \\
\hline
`Natural' inflation & $1 + \cos(\phi/f)$ & any $n<1$ & 0 \\
\hline
Hybrid inflation (standard) & $1 + B\phi^2$ & 1 & 0 \\
Hybrid inflation (extreme) & $1 + B\phi^2$ & $1 < n < 1.15$ & $\sim 0$ \\
\hline
\end{tabular}
\caption[pred]{\label{arl:pred} The spectral index and gravitational wave 
contribution for a range of inflation models.}
\end{table}

Observations have some way to go before the power-law approximation becomes 
inadequate. Consequently ...
\begin{itemize}
\item Slow-roll inflation adds two, and only two, new parameters to 
large-scale structure.
\item Although $\epsilon$ and $\eta$ are the fundamental parameters, it is 
best to take them as $n$ and $R$.
\item Inflation models predict a wide range of values for these. Hence 
inflation makes no definite prediction for large-scale structure.
\item However, this means that large-scale structure observations, and 
especially microwave background observations, can strongly discriminate 
between inflationary models. When they are made, most existing inflation 
models will be ruled out.
\end{itemize}

\subsection{Testing the idea of inflation}

The moral of the previous section was that different inflation models
lead  to very different models of structure formation, spanning a wide
range of  possibilities. That means, for example, that a definite
measure of say the  spectral index $n$ would rule out most inflation
models. But it would always  be possible to find models which did give
that value of $n$. Is there any  way to try and test the idea of
inflation, independently of the model  chosen?

The answer, in principle, is yes. In the previous section we introduced 
three observables (in addition to the overall normalization), namely $n$, 
$R$ and $n_{{\rm G}}$. However, they depend only on two fundamental 
parameters, namely $\epsilon$ and $\eta$.\cite{LL} We can therefore 
eliminate $\epsilon$ and $\eta$ to obtain a relation between observables, 
the {\em consistency equation}
\begin{equation}
R = - 2 \pi n_{{\rm G}} \,.
\end{equation}
This relation has been much discussed in the literature.\cite{recon,LLKCBA}
It is {\em independent} of the choice of inflationary model (though it does 
rely on the slow-roll and power-law approximations).

The idea of a consistency equation is in fact very general. The point is
that we have obtained two continuous functions, $\delh(k)$ and $A_{{\rm
G}}(k)$, from a single continuous function $V(\phi)$. This can only be 
possible if the functions $\delh(k)$ and $A_{{\rm G}}(k)$ are related,
and  the equation quoted above is the simplest manifestation of such a
relation.

Vindication of the consistency equation would be a remarkably convincing
test of the inflationary paradigm, as it would be highly unlikely that
any other production mechanism could entangle the two spectra in the way
inflation does. Unfortunately though, measuring $n_{{\rm G}}$ is a much
more challenging observational task than measuring $n$ or $R$ and is
likely to be
beyond even next generation observations. Indeed, this is a good point
to remind the reader that even if inflation is right, only one model can
be right and it is perfectly possible (and maybe even probable, see
Ref.~\cite{lythnew}) that that model has a very low amplitude of
gravitational waves and that they will never be detected.

\section{The inflationary origin of structure}

At the summer school where these lectures were given, models of
structure formation were described in detail by Joe Silk and for a
detailed treatment I refer you to his corresponding article. Here I will
address those issues of direct relevance to the inflationary cosmology.

\subsection{The parameters}

The initial goal of structure formation studies is to accurately
determine the fundamental parameters describing our Universe. So far
I've stressed the three inflationary parameters, $\delta_{{\rm H}}$, $n$
and $r$, which describe the initial perturbations which inflation
generates. However, except on very large scales where they remain
untouched by causal processes, we do not see the original perturbations
but rather than perturbations after they have been processed by a
variety of physical mechanisms. This processing depends on many
quantities, all of which must be either fixed by assumption or determined
from observations. A basic list features four categories; the global
dynamics, the way in which the matter content is divided amongst the
different particle species, astrophysics effects such as reionization
which would affect the microwave background photons, and the initial
perturbation spectrum that we are here assuming comes from inflation. A
possible list might look like this
\begin{enumerate}
\item {\bf Global dynamics}
\begin{description}
\item Hubble constant $h$ \hspace{1cm} $*$
\item Spatial curvature $k$
\end{description}
\item {\bf Matter content}
\begin{description}
\item Baryons $\Omega_{{\rm B}}$ \hspace*{1cm} $*$
\item Hot dark matter? $\Omega_{{\rm HDM}}$
\item Cosmological constant? $\Lambda$ \hspace{1cm} $(*)$
\item Massless species? $g_*$
\end{description}
\item {\bf Astrophysics}
\begin{description}
\item Reionization optical depth $\tau$ \hspace*{1cm} $*$
\end{description}
\item {\bf Initial perturbations}
\begin{description}
\item Amplitude $\delta_{{\rm H}}(k= a_0 H_0)$ \hspace*{1cm} $*$
\item Spectral index $n$ \hspace*{1cm} $*$
\item Gravitational waves $r$ 
\end{description}
\end{enumerate}
A cold dark matter contribution is not mentioned under matter content as
it is assumed to take the value required to make the sums add up
(i.e.~to give the right spatial curvature $k$ given the other matter
densities). 

In this list, I've starred those parameters which need to be included in
even the most minimal model, while the rest can be set to some
particular value by assumption. I've partially starred the cosmological
constant because although most people would like to set it to zero, the
observational case for a non-zero value is near to overwhelming.

\subsection{The inflationary energy scale}

The most solid observational result is the interpretation of the cosmic
microwave anisotropies seen by COBE as giving the amplitude of the
initial power spectrum. COBE is a particularly powerful probe because
its large beam size makes it sensitive only to scales much larger than
the horizon size when the microwave background formed. The perturbations
are therefore seen in their primordial form, and depend only
on the initial perturbations and not all the other
parameters.~\footnote{There is a residual dependence on $\Omega_0$ and
$\Lambda$ which determine the relation between the metric perturbations
and the matter perturbations, and also the evolution of perturbations,
but that is easily dealt with. I will assume critical density for
simplicity.} 

The COBE normalization requires the perturbation at the present Hubble
scale, $\delta_{{\rm H}} \equiv \delta_{{\rm H}}(k=a_0 H_0)$, to be
given by~\cite{BLW}
\begin{equation}
\delta_{{\rm H}} \simeq 2 \times 10^{-5} \,.
\end{equation}
Since
\begin{equation}
\delta_{{\rm H}}^2 = \frac{32}{75} \, \frac{V}{m_{{\rm Pl}}^4} \, 
    \frac{1}{\epsilon} \,,
\end{equation}
then unless $\epsilon$ proves to be tiny (say much less than a
hundredth) this will give
\begin{equation}
V^{1/4} \simeq 10^{-3} m_{{\rm Pl}} \simeq 10^{16} \, {\rm GeV} \,,
\end{equation}
at the time when observable scales crossed outside the horizon, pretty
much the scale that particle physicists associate with Grand Unified
Theories. 

\subsection{Beyond the energy scale}

To go beyond the energy scale entails bringing together as wide a range
of observations as possible to try and constrain the wide parameter
family. When restricted parameter sets are considered quite interesting
constraints can be quoted, but these weaken once the parameter space is
widened. Until recently no-one attempted a plausibly large parameter
space, but recently Tegmark~\cite{Teg9} considered a nine-parameter
family of models, including the three inflationary parameters, which is
the first attempt to get to grips with the large families of models that
need to be considered for us to become convinced we are on the right
track. 

At present, observations are only quite weakly constraining concerning
quantities beyond the inflationary energy scale. The spectral index is
known to lie near one, with the plausible range, depending on what
parameters one allows to vary, stretching from perhaps 0.8 to 1.2. As it
happens, that is more or less the range which current inflation models
tend to cover, and so most models survive. The holy grail for inflation
model building is an accurate measurement of $n$, say with an error bar
of around 0.01 or better. Such a measurement would exclude the vast
majority of the models currently under discussion. MAP, and certainly
{\sc Planck}, ought to be able to deliver a measurement at around this
accuracy level, and perhaps may even be able to see deviations from
perfect power-law behaviour.\cite{KosT,CGL}

At the moment there is no evidence favouring a gravitational wave
contribution to COBE, but equally the upper limit on such a
contribution, perhaps around $r < 1$ depending on other parameters (see
Ref.~\cite{Zetal} for a recent analysis), is
unable to rule out much in the way of interesting models (though it is a
combination of the constraints on $n$ and $r$ that kills extended
inflation). If such a contribution can be identified, it will be very
strong support for inflation, but since many models, especially of the
currently-popular hybrid type, predict insignificant gravitational wave
production, even the strongest achievable upper limits may tell us
nothing. 

A particularly powerful test of inflation will be whether or not the
microwave anisotropy spectrum (the $C_\ell$) proves to contain an
oscillatory peak structure.\cite{HW} Such a structure is evidence of
phase coherence in the evolution of perturbations (meaning that the
perturbations of a given wavenumber are at a calculable phase of
oscillation). Such phase coherence would indicate that perturbations are
entirely in the growing mode, which in turn implies that they have been
evolving sufficiently long for the decaying mode to become negligible.
For modes around the horizon scale at decoupling, this implies that they
were already in place while well outside the horizon, which is a
characteristic of inflationary perturbations (a characteristic not
shared by topological defect models, for instance). This fairly
qualitative test, if satisfied, will provide strong support for the
inflationary paradigm, while if a multiple peak structure is not
observed that will imply that the inflationary mechanism is not the sole
source of perturbations in the Universe.

\section{Summary}

In this article I have introduced some of the facets of inflation in a
fairly simple manner. If you are interested in going beyond this, then
the inflationary production of perturbations is reviewed in
Ref.~\cite{LLKCBA}, inflation and structure formation in
Ref.~\cite{LLrep} and particle physics aspects of inflation in
Ref.~\cite{LR}. 

At present, inflation is the most promising candidate theory for the
origin of perturbations in the Universe. Different inflation models lead
to discernibly different predictions for these perturbations, and hence
high-accuracy measurements are able to distinguish between models,
excluding either all or the vast majority of them. 

Since its inception, the inflationary cosmology has been a gallery of
different models, and the gallery has continually needed extension after
extension to house new acquisitions. In all the time up to the present,
very few models have been discarded. However, the near future holds
great promise to finally begin to throw out inferior models, and, if the
inflationary cosmology survives as our model for the origin of
structure, we can hope to be left with only a narrow range of models to
choose between.

\section*{Acknowledgments}

The author was supported in part by the Royal Society.




\begin{thebibliography}{99}
\bibitem{Guth} A. H. Guth, Phys. Rev. {\bf 23}, 347 (1981).
\bibitem{LLrep} A. R. Liddle and D. H. Lyth, Phys. Rep {\bf 231}, 1 (1993).
\bibitem{LL} A. R. Liddle and D. H. Lyth, Phys. Lett. B {\bf 291}, 
	391 (1992).
\bibitem{SB} D. S. Salopek and J. R. Bond, Phys. Rev. D {\bf 42}, 3936
	(1990).
\bibitem{LPB} A. R. Liddle, P. Parsons and J. D. Barrow, Phys. Rev. D 
	{\bf 50}, 7222 (1994).
\bibitem{Linbook} A. D. Linde, {\em Particle Physics and Inflationary
	Cosmology}, Harwood Academic, Chur, Switzerland (1990).
\bibitem{KT} E. W. Kolb and M. S. Turner, {\em The Early Universe}, 
	Addison-Wesley, Redwood City, California (1990) 
	[updated paperback edition 1994].
\bibitem{KLS} L. Kofman, A. D. Linde and A. A. Starobinsky, Phys. Rev.
	Lett. {\bf 73}, 3195 (1994); A. D. Linde, astro-ph/9601004; L.
	Kofman, astro-ph/9605155.
\bibitem{preheat} Y. Shtanov, J. Traschen and R. Brandenberger, Phys. Rev.
	D {\bf 51}, 5438 (1995); D. Boyanovsky, M. D'Attanasio, H. de
	Vega, R. Holman, D.-S. Lee and A. Singh, Phys. Rev. D {\bf 52},
	6805 (1995).
\bibitem{LR} D. H. Lyth and A. Riotto, to appear, Phys. Rep.,
    hep-ph/9807278.  
\bibitem{MAP} {\sc map} home page at {\tt http://map.gsfc.nasa.gov/}.
\bibitem{planck} {\sc Planck} home page at {\tt
	http://astro.estec.esa.nl/Planck/}.
\bibitem{hybrid} A. D. Linde, Phys. Lett. B {\bf 259}, 38 (1991), Phys. 
	Rev. D {\bf 49}, 748 (1994); E. J. Copeland, A. R. Liddle, D. H.
	Lyth, E. D. Stewart and D. Wands, Phys. Rev. D {\bf 49}, 6410
	(1994).
\bibitem{W93} C. M. Will, {\em Theory and Experiment in Gravitational
	Physics}, Cambridge University Press (1993).
\bibitem{conform} B. Whitt, Phys. Lett. {\bf 145B}, 176 (1984);
	K. Maeda, Phys. Rev. D {\bf 39}, 3159 (1989); D. Wands, Class.
	Quant. Grav. {\bf 11}, 269 (1994).
\bibitem{extinf} D. La and P. J. Steinhardt, Phys. Rev. Lett. {\bf 62}, 376
	(1989); E. W. Kolb, Physica Scripta {\bf T36}, 199 (1991).
\bibitem{open} J. R. Gott, Nature {\bf 295}, 304 (1982); M. Sasaki, T.
	Tanaka, K. Yamamoto and J. Yokoyama, Phys. Lett. B {\bf 317}, 
	510 (1993); M. Bucher, A. S. Goldhaber and N. Turok, Phys. Rev. D 
	{\bf 52}, 3314; A. D. Linde and A. Mezhlumian, Phys. Rev. D {\bf
	52}, 6789 (1995).
\bibitem{HT} S. W. Hawking and N. Turok, Phys. Lett. B {\bf 425}, 25
    (1998); A. D. Linde, Phys. Rev. D {\bf 58}, 083514 (1998).
\bibitem{COBE1} G. F. Smoot et al., Astrophys. J. {\bf 396}, L1 (1992).
\bibitem{COBE4} C. L. Bennett et al., Astrophys. J. {\bf 464}, L1 (1996).
\bibitem{LLKCBA} J. E. Lidsey, A. R. Liddle, E. W. Kolb, E. J. Copeland,
	T. Barriero and M. Abney, Rev. Mod. Phys {\bf 69}, 373 (1997).
\bibitem{GL} I. J. Grivell and A. R. Liddle, Phys. Rev. D {\bf 54}, 7191
    (1996).
\bibitem{SL} E. D. Stewart and D. H. Lyth, Phys. Lett. B {\bf 302}, 171
    (1993).
\bibitem{GW} J. Garc\'{\i}a-Bellido and D. Wands, Phys. Rev. D {\bf
    54}, 7181 (1996).
\bibitem{BLW} E. F. Bunn and M. White, Astrophys. J. {\bf 480}, 6 (1987);
	E. F. Bunn, A. R. Liddle and M. White, Phys. Rev. D 
	{\bf 54}, 5917R (1996).
\bibitem{recon} E. J. Copeland, E. W. Kolb, A. R. Liddle and J. E. Lidsey,
	Phys. Rev. D {\bf 48}, 2529 (1993), {\bf 49}, 1840 (1994).
\bibitem{lythnew} D. H. Lyth, Phys. Rev. Lett. {\bf 78}, 1861 (1997).
\bibitem{Teg9} M. Tegmark, preprint {\tt astro-ph/9809201}.
\bibitem{KosT} A. Kosowsky and M. Turner, Phys. Rev. D {\bf 52}, 1739
    (1995).
\bibitem{CGL} E. J. Copeland, I. J. Grivell and A. R. Liddle, Mon. Not.
    R. Astron. Soc. {\bf 298}, 1233 (1998).
\bibitem{Zetal} J. Zibin, D. Scott and M. White, preprint {\tt
    astro-ph/9901028}. 
\bibitem{HW} W. Hu and M. White, Phys. Rev. Lett. {\bf 77}, 1687 (1996).
\end{thebibliography}
\end{document}